\documentclass[useAMS,usenatbib]{mn2e}
\usepackage{graphics,rotate}
\def\gtsim{\mathrel{\hbox{\rlap{\hbox{\lower4pt\hbox{$\sim$}}}\hbox{$>$}}}}
\def\lesssim{\mathrel{\hbox{\rlap{\hbox{\lower4pt\hbox{$\sim$}}}\hbox{$<$}}}}

\def\etal{{\it et al.\ }}
\def\eg{{\it e.g.\ }}

\def\spose#1{\hbox to 0pt{#1\hss}}
\def\approxlt{\mathrel{\spose{\lower 3pt\hbox{$\sim$}}
        \raise 2.0pt\hbox{$<$}}}
\def\approxgt{\mathrel{\spose{\lower 3pt\hbox{$\sim$}}
        \raise 2.0pt\hbox{$>$}}}
\def\approxpropto{\mathrel{\spose{\lower 3pt\hbox{$\sim$}}
        \raise 2.0pt\hbox{$\propto$}}}
\mathchardef\twiddle="2218

\def\multleft#1{\hbox to size{\vbox {\halign {\lft{##}\cr #1}}\hfill}\par}
\def\multright#1{\hbox to size{\vbox {\halign {\rt{##}\cr #1}}\hfill}\par}

\def\today{\ifcase\month\or January\or February\or March\or April\or May\or
      June\or July\or August\or September\or October\or November\or December\fi
      \space\number\day, \number\year}
\def\<{\thinspace}

\def\km{{\rm\thinspace km}}

\def\Mpc{{\rm\thinspace Mpc}}

\def\s{{\rm\thinspace s}}

\def\kmps{\hbox{$\km\s^{-1}\,$}}

\def\kmpspMpc{\hbox{$\kmps\Mpc^{-1}$}}

\title[Measuring $H_0$ using X-ray and SZ observations of galaxy clusters]{An improved approach to 
measuring $H_0$ using X-ray and SZ observations of galaxy clusters}

\author[Schmidt, Allen \& Fabian]
        {R. W. Schmidt$^1$\thanks{E-mail: rschmidt@astro.physik.uni-potsdam.de},
        S. W. Allen$^2$ and A.C. Fabian$^2$\\ 
        1. Institut f\"ur Physik, Universit\"at Potsdam, Am Neuen
        Palais 10, 14469 Potsdam, Germany\\
        2. Institute of Astronomy, University of Cambridge, Madingley Road,
        Cambridge CB3 0HA}
\date{Draft \today
        }

\pagerange{\pageref{firstpage}--\pageref{lastpage}}
\pubyear{}

\begin{document}

\maketitle

\label{firstpage}

\begin{abstract}
\noindent 
We present an improved method for predicting the Sunyaev-Zeldovich
(SZ) effect in galaxy clusters from spatially-resolved, spectroscopic 
X-ray data. Using the deprojected electron density and temperature
 profiles measured within a fraction of the virial radius, and assuming 
a \citet*{Navarro95} mass model, we
show how the pressure profile of the X-ray gas  can be extrapolated to
large radii, allowing the  Comptonization parameter profile for the cluster
to be predicted precisely. We apply our method to  Chandra
observations of three X-ray luminous,  dynamically relaxed
clusters with published SZ data:  RX J1347.5-1145, Abell 1835 and
Abell 478.  Combining the predicted and observed SZ signals, we
determine improved estimates for the Hubble constant from each cluster
and obtain a weighted mean of $H_0=69\pm8$\kmpspMpc~for a
cosmology with $\Omega_{\rm m}=0.3$ and $\Omega_{\Lambda}=0.7$. This 
result is in good agreement with independent findings from the 
Hubble Key Project and the combination of cosmic microwave background
and galaxy cluster data.
\end{abstract}


\section{Introduction}
\label{intro}

The inverse Compton scattering of cosmic microwave background (CMB)
photons by hot electrons in galaxy clusters leads to a distortion of
the CMB spectrum along the line of sight, known as the
Sunyaev-Zeldovich (SZ, \citealt{Sunyaev72}) effect. The magnitude of
the SZ effect is determined by the Comptonization parameter of the
cluster gas, $y(r)$, which is proportional to the line of sight
integral of the gas pressure.

It was recognized swiftly \citep*{Silk78,Cavaliere79} that for an
assumed, simple geometry, the combination of X-ray and SZ observations
can be used to measure the angular diameter distance to a cluster. The
ratio of the observed (based on radio/sub-mm observations) and
predicted (based on X-ray observations) SZ signals is proportional to
the square of the angular diameter distance, making this, potentially,
an exceptionally powerful technique for probing the cosmic distance
scale.

Although the SZ effect is now employed frequently to determine
extragalactic distances \citep*[e.g.,
][]{Mauskopf00,Carlstrom01,Mason01,Jones01,Reese02}, most studies to
date have relied on the application of the $\beta$-model
\citep{Cavaliere78} in their X-ray analyses. The $\beta$-model
provides a simple analytical approximation to the spatial distribution
of the cluster gas. These studies have also generally assumed 
isothermality and relied on broad beam measurements of the mean, 
emission-weighted temperature when calculating the
gas pressure.

With the {Chandra} X-ray Observatory and XMM-Newton it is now possible
to measure the temperature, density and pressure profiles of the
cluster gas precisely. The surface brightness distribution, in
particular, can be resolved on sub-arcsecond scales. Chandra
observations have shown that $\beta$-models do 
not, generally, provide good descriptions of the X-ray gas distributions 
in the cores of dynamically-relaxed clusters, and that the gas 
temperature is not isothermal but drops sharply within 
cluster cores \citep[e.g., ][]{Schmidt01,Allen01b,Kaastra04}.  
The main limitation of the new X-ray observations is that (due 
to background levels and restricted fields of view; $8\times8$ 
arcmin$^2$ for the Chandra ACIS-S detector) 
the gas temperature can only be measured directly out to 
radii $r \approxlt$ one third of the virial radius for most clusters. 
Since predictions of the SZ effect require the line-of-sight pressure 
integral through the cluster, extrapolation of the 
X-ray results is therefore required for combined X-ray/SZ work.

In this paper we present a recipe for calculating the predicted 
SZ effect in clusters which makes use of the spatially-resolved
spectroscopic techniques described in our earlier work, and which 
includes a new method for extrapolating the X-ray pressure profiles 
beyond the directly observed region. Analytical formulae for the 
electron density and temperature profiles of the X-ray gas at 
large radii are obtained under the assumption of hydrostatic 
equilibrium. These can be computed easily and used for SZ analysis.
We show that our method, which takes full account of the 
observed density and temperature profiles of the X-ray emitting gas, 
provides a significant improvement in accuracy with respect to the 
standard isothermal $\beta$-model.

The outline of this paper is as follows: In Section 2 we review the 
standard $\beta$-model approach and describe our improved method. 
In Section 3 we apply our method to Chandra X-ray and SZ observations 
of three highly X-ray luminous, dynamically relaxed galaxy clusters: RX
J1347.5-1145 ($z=0.451$), Abell 1835 ($z=0.2523$) and Abell 478
($z=0.088$). In Section 4 we use the combined data to measure the Hubble 
constant. In Section 5 we compare our results to those obtained 
using the standard isothermal $\beta$-model approach. 
A summary of the results can be found in Sect.~\ref{conclusions}.

Unless stated otherwise, all quantities are are quoted  for  a Hubble
constant $H_0=70\kmpspMpc$, matter density $\Omega_{\rm m}=0.3$ and
vacuum energy density $\Omega_{\Lambda}=0.7$. Error  bars correspond
to 1\,$\sigma$ (68.3\%) confidence.

\section{Predicting the SZ effect}
\subsection{The standard $\beta$-model approach}
\label{betamodelsect}

Within the context of the isothermal $\beta$-model
\citep{Cavaliere78}, the surface brightness profile of a galaxy
cluster can be written as
\begin{equation}
I(\theta)\propto \left[1+\left(\frac{\theta}{\theta_{\rm c}}\right)^2
\right]^{-3\,\beta+\frac{1}{2}},
\label{betasurface}
\end{equation}
where $\theta_{\rm c}$ is the angular core radius and $\beta$ is the
slope parameter. By calculating the X-ray emission due to hot cluster
gas at a temperature $T$ \citep[e.g., ][]{Kaastra93,Liedahl95}, one can
invert this to yield the central electron density $n_{{\rm e}\,0}$ of
the corresponding intrinsic (3-dimensional) electron density profile
\begin{equation}
n_{\rm e}(r)=n_{{\rm e}\,0} \left[1+\left(\frac{r}{r_{\rm c}}\right)^2
\right]^{-\frac{3\,\beta}{2}},
\label{betamodel}
\end{equation}
where the core radius $r_{\rm c}$ follows from the angular diameter
distance $d$ of the cluster, $r_{\rm c}=d\times\theta_{\rm c}$.

The Comptonization parameter along a line of sight at an impact
parameter $R$ is defined by
\begin{equation}
y(R)=\frac{2\,\sigma_{\rm T}\,k_{\rm B}}{m_{\rm e}\,c^2}\,\int_R^{\infty}\,
\frac{n_{\rm e}(r)\,T(r)}{\sqrt{r^2-R^2}}\,r\,{\rm d}\,r
\label{y-integral}
\end{equation}
\citep[e.g., ][ where we have converted the equation into an integral
along radius $r$]{Birkinshaw99}. Here $\sigma_{\rm T}$ is the Thomson
cross-section, $k_{\rm B}$ the Boltzmann constant, $c$ the speed of
light, and $m_{\rm e}$ the electron mass.  Eq.~(\ref{y-integral}) can
be integrated analytically for the isothermal $\beta$-model
\citep[e.g., ][]{Birkinshaw99} to give
\begin{equation}
y(R)=y_0 \left[1+\left(\frac{R}{r_{\rm
      c}}\right)^2\right]^{-\frac{3}{2}\beta+\frac{1}{2}},
\label{betacomp}
\end{equation}
with the normalization constant
\begin{equation}
y_0=n_{{\rm e}\,0}\, r_{\rm c}\, \sigma_{\rm T} \left(\frac{k_{\rm B} T}
{m_{\rm e} c^2}\right) B(\frac{1}{2},\frac{3\beta}{2}-\frac{1}{2})
\label{y0}
\end{equation}
\citep{Mauskopf00}, where $B(a,b)=\Gamma(a) \Gamma(b)/\Gamma(a+b)$
is the beta function.

In previous work, the temperature $T$ has usually been taken to be the 
mean emission-weighted
temperature of the X-ray gas, determined with broad-beam instruments.
However, as discussed in Section 1, observations with the Chandra
and XMM-Newton satellites have shown that regular, dynamically relaxed
galaxy clusters are not isothermal. This temperature variation 
should be accounted for in
the analysis.\footnote{Uncertainties in the determination of the distance
scale associated with the assumption of isothermality were
previously discussed by, e.g., \citet*{Inagaki95} and \citet{Yoshikawa98}.}

Finally, clusters are not infinitely large, as is usually assumed when
calculating the predicted Comptonization parameter profile using
the $\beta$-model. In what follows we show how modern X-ray data 
can be used to integrate eq.~(\ref{y-integral}) without the 
assumption of isothermality, taking account of 
the finite size of galaxy clusters. In particular, where the region 
of the cluster directly observed in X-rays is small, the use of an 
extrapolation 
procedure like the one described here can become important.

\subsection{An improved approach}
\label{method}

\subsubsection{Discretization of the $y$-parameter calculation}
\label{step1}

In order to calculate the expected y-parameter profile
(eq.~\ref{y-integral}) for a particular galaxy cluster, the
temperature profile $T(r)$ and the electron density profile $n_{\rm
e}(r)$ need to be known. In practise, deprojection analyses of 
Chandra or XMM-Newton data 
(e.g., \citealt{Schmidt01} and references therein)
provide the electron density and X-ray gas
temperature in $N$ discrete shells, with inner and outer radii $r_{{\rm
in},i}$ and $r_{{\rm out},i}$, respectively.
We can thus recast
eq.~(\ref{y-integral}) into a sum starting with shell j at 
impact parameter $R$:
\begin{eqnarray}
y(R)&=&\frac{2\,\sigma_{\rm T}\,k_{\rm B}}{m_{\rm
e}\,c^2}\,\left(\,n_{{\rm e},{\rm j}}\,T_{\rm j}\,\sqrt{r_{{\rm
out},{\rm j}}^2-R^2}\right. \nonumber\\
&&\left.+\sum_{i={\rm j}+1}^{N}\,n_{{\rm e},i}\,T_i\,\left(\sqrt{r_{{\rm
out},i}^2-R^2}- \sqrt{r_{{\rm in},i}^2-R^2}\right)\right).
\label{y-sum}
\end{eqnarray}
For clusters with temperatures above about 8 keV, relativistic
corrections also become significant. Analytic and fitting relations for the
relativistic corrections to the SZ effect have been worked out by
several groups: \citet{Itoh98,Challinor98,Sazonov98}. Using the
analytical formulae of \citet{Challinor98} up to second order, we 
determine the relativistic correction factor $\chi$ that
needs to be applied to each summand of~(\ref{y-sum}), so that $\Delta
y_i\rightarrow\chi\,(T_i)\times \Delta y_i$.

\subsubsection{Extrapolation of the temperature and gas density
  profiles}
\label{extrap_proc}

In the case of the {Chandra} X-ray Observatory, the detector size and
the particle background limit the regions of clusters for which direct 
temperature measurements can be made ($r \leq r_1$) to a fraction 
of the virial
radius. In order to calculate the predicted Comptonization parameter precisely,
we need to extrapolate the temperature and electron density profiles
out to the edge of the cluster, $r_2$ - or at least past the point where 
significant contributions to $y(r)$ are made. Here, for
convenience, we will set $r_1=r_{2500}$ and $r_2=r_{200}$ 
(corresponding to the radii within which the mean 
enclosed mass density is $\Delta=2500$ and $\Delta=200$ times the 
critical density
of the Universe $\rho_{\rm crit}(z)$ at redshift $z$,
respectively). $r_{2500}$ is typical of the 
outer radii for which useful information on the gas 
temperature profile can be obtained from Chandra observations. 
$r_{200}$ is used to mark the outer edge of the cluster. 
(In Sect.~\ref{conclusions} we shall show that the precise choice 
of $r_2$ does not affect the results significantly.)

We parametrize the total mass distributions in 
the clusters using an NFW model
\begin{equation}
\rho(r) = {{\rho_{\rm crit}(z)\,\delta_{\rm c}} \over {  ({r/r_{\rm s}}) 
\left(1+{r/r_{\rm s}} \right)^2}},
\label{nfw}
\end{equation}
where $\rho(r)$ is the mass density, $r_{\rm s}$ is the scale radius, 
$\rho_{\rm crit} =
\frac{3H(z)^2}{8 \pi G}$, $H(z)$ is the Hubble constant, 
G is the
gravitational constant, $\delta_{\rm c} = \frac{200}{3} \frac{
c^3}{ {{\rm ln}(1+c)-{c/(1+c)}}},$ and $c$ is the
concentration parameter with $c=r_{200}/r_{\rm s}$.

The temperature
$T({r_1})$, electron density $n_{\rm e}(r_1)$, and 
gas mass-weighted temperature $T_{{\rm m},r_1}$ within $r_1$ are
assumed to be known from direct spatially-resolved spectroscopy. 
We extrapolate the electron density from $r_1$ 
to $r_2$ using a power-law model 
$n_{\rm e}(r)\sim r^{-\gamma}$. The hydrostatic equation is then used 
to determine the temperature
solution $T(r)$ for that power-law electron density profile in the given 
NFW potential (see Appendix~\ref{extrap}). 

We fit the exponent $\gamma$ of the
electron density profile so that the model complies with two fixed quantities:
\begin{enumerate}
\item the observed temperature $T({r_1})$ at $r_1$
\item the {\em gas-mass} weighted temperature $T_{{\rm m},r_2}$ at
$r_2$.
\end{enumerate}
The
power-law model approximates the true electron density 
profile between $r_1$ and $r_2$. (It is assumed that the galaxy cluster ends at $r_2$.) To determine the gas-mass weighted temperature $T_{{\rm m},r_2}$
within $r_2$, we use the \citet{Lokas01} solution for the radial
dependence of the ratio of kinetic energy $W_{\rm kin}$ and potential
energy $W_{\rm pot}$ in an NFW potential with an isotropic velocity
dispersion. The key assumption here is that the gas-mass
weighted temperature $T_{{\rm m},r}$ inside a radius $r$ is
proportional to the kinetic energy of the mass distribution inside
this radius (e.g., \citealt{Eke98}). We then calculate the change,
$q$, of the ratio $W_{\rm
kin}/W_{\rm pot}$ between $r_1$ and $r_2$ as
\begin{equation}
q=\frac{W_{\rm kin}(r_1)/W_{\rm pot}(r_1)}{W_{\rm kin}(r_2)/W_{\rm
    pot}(r_2)}=\frac{T_{{\rm m},r_1}/\left[M_{\rm
    tot}(r_1)/r_1 \right]}{T_{{\rm m},r_2}/\left[M_{\rm
    tot}(r_2)/r_2 \right]},
\label{lokas}
\end{equation}
where $M_{\rm tot}(r)$ is the total enclosed mass within radius
$r$. The quantity $q$ can be computed 
as a function of the scale radius $r_{\rm s}$ and the
concentration parameter $c$ of the NFW profile using
the analytical formulae included
in Appendix~\ref{lokmam}.

Thus, given both the gas-mass weighted temperature $T_{{\rm
    m},r_1}$ at $r_1$ and a specific NFW mass model from the 
Chandra data, the gas-mass
    weighted temperature $T_{{\rm m},r}$ at $r_2$ can be obtained from
    eq.~(\ref{lokas}). For the correct electron density slope
    $\gamma$, this is equal to the temperature that follows from the
    temperature solution~(eq.~\ref{tprof_analyt}) and gas density
    profile $\rho_{\rm g}(r)$ (for which we assume $\rho_{\rm
    g}(r)=1.1345\,m_{\rm p}\,n_{\rm e}(r)$) via the integral
\begin{equation}
T_{{\rm m},r_2}=\frac{1}{M_{{\rm g},r_2}} \left( M_{{\rm g},r_1}
T_{{\rm m},r_1} + 4\pi \int_{r_1}^{r_2} T(r) \rho_{\rm g}(r) r^2 {\rm d}r
\right),
\label{intconst}
\end{equation}
where $M_{{\rm g},r}$ is the gas mass inside the radius $r$.

We note that in cases where the X-ray data only extend to radii 
$r_0 \ll r_{2500}$ (within which cooling and/or heating effects 
may have modified $T_{{\rm m},r_0}$ significantly), one should 
extrapolate the gas density (Appendix~\ref{extrap}) from 
$r_0$ to $r_{200}$, use this extrapolation to estimate the 
gas temperature and density at $r_1=r_{2500}$, and then apply 
eq.~(\ref{lokas}) between $r_{2500}$ and $r_{200}$ as usual. 
(This is possible because 
the extrapolation recipe is attached continuously to the Chandra data 
so that $T_{{\rm m},r}$ can be calculated for any exponent $\gamma$ of 
the electron density extrapolation, regardless of where we attach the 
extrapolation.) We show below that this approach leads to robust 
answers in the case of Chandra observations of Abell 478, where the 
data cover only a relatively small radial range.

\begin{table}
\caption{Details of the {Chandra} NFW mass models}
\label{tab:nfw}
\begin{tabular}{llcc}
Object & redshift & $c$ & $r_{\rm s}$ (Mpc) \\
\hline
RX J1347.5-1145 & 0.451 & $6.34^{+1.61}_{-1.36}$ & $0.37^{+0.18}_{-0.12}$ \\
Abell\,1835     & 0.2523 & $4.21^{+0.53}_{-0.61}$ & $0.55^{+0.18}_{-0.09}$ \\
Abell\,478      & 0.088 & $3.88^{+0.28}_{-0.36}$ & $0.61^{+0.12}_{-0.07}$ 
\end{tabular}
\end{table}

\section{Application to published Chandra and SZ observations}
\label{observations}

In this section we apply our method to Chandra and SZ 
observations of three
galaxy clusters: RXJ1347.5-1145, Abell 1835 and Abell 478. 
The X-ray data were
originally published by \citet{Allen02}, \citet{Schmidt01} and
\citet{Sun03}, respectively.

\subsection{Analysis of the X-ray data}

\begin{figure*}
\resizebox{0.7\textwidth}{!}{\includegraphics{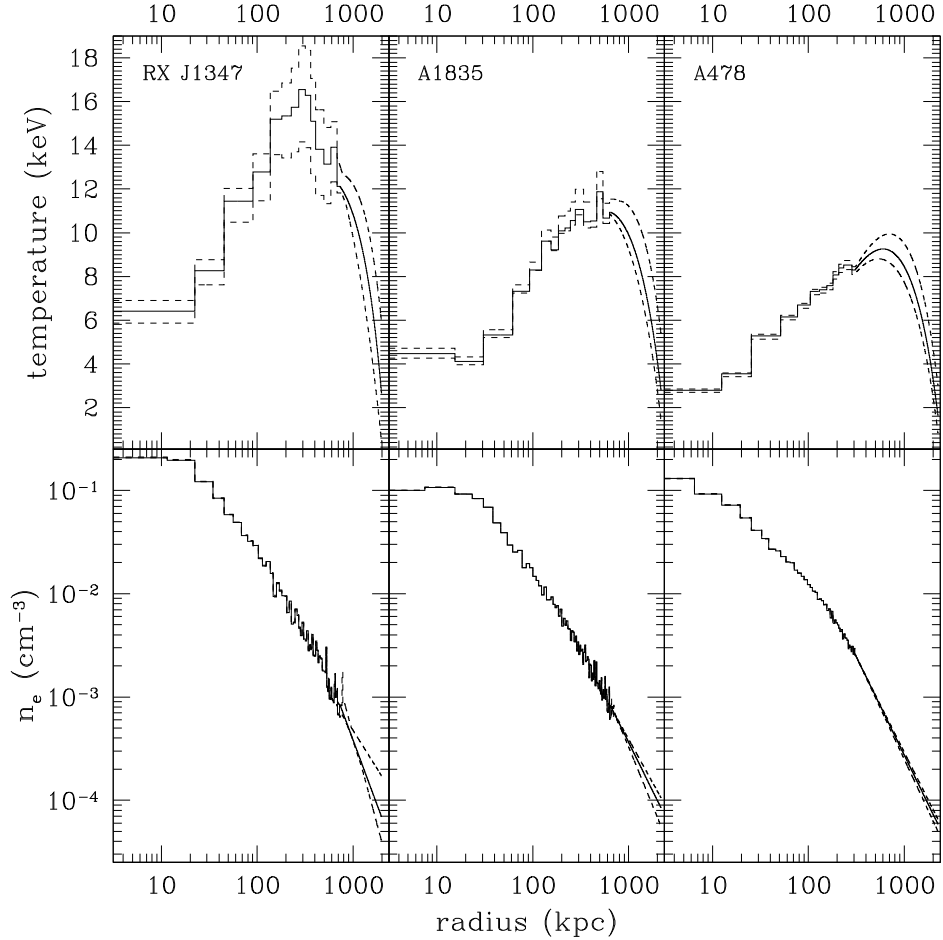}}
\caption{Extrapolated temperature and electron density profiles for RX
  J1347.5-1145, Abell 1835 and Abell 478. The best fit profiles are
  indicated by the solid lines. The upper and lower envelopes of the 
  model ranges are plotted with dashed lines. For clarity we have only
  plotted the profiles up to the smallest $\Delta=200$ overdensity
  radius of each model ensemble (see Tab.~\protect\ref{tab:extrap}).}
\label{profiles}
\end{figure*}

\begin{table*}
\caption{Details of the cluster model extrapolation}
\label{tab:extrap}
\begin{tabular}{lccccccc}
Object & $r_1$ & $r_2$ & $T({r_1})$ & $T_{{\rm m},r_2}$ & n$_{\rm e}(r_1)$ &
q & $\gamma$\\
&(Mpc) & (Mpc) & (keV) & (keV) & $10^{-4}$cm$^{-3}$\\
\hline
RX J1347.5-1145 & 0.73$^{+0.08}_{-0.09}$ & 2.34$^{+0.40}_{-0.35}$ &
12.13$^{+1.46}_{-0.41}$ & $9.69^{+1.18}_{-1.52}$ & $8.44^{+0.82}_{-0.20}$ &
1.25$^{+0.06}_{-0.05}$ & 2.50$^{+0.97}_{-1.02}$\\
Abell 1835 & 0.67$^{+0.05}_{-0.03}$ & 2.36$^{+0.26}_{-0.18}$ &
11.34$^{+0.20}_{-0.49}$ & 7.97$^{+0.75}_{-0.63}$ & 7.46$^{+0.81}_{-0.47}$ &
1.36$^{+0.05}_{-0.04}$ & 1.88$^{+0.39}_{-0.20}$\\
Abell 478 & $0.66^{+0.03}_{-0.02}$ & $2.36^{+0.18}_{-0.11}$ & 
$8.29^{+0.16}_{-0.21}$ & $6.41^{+0.48}_{-0.32}$ & $5.92^{+0.60}_{-0.77}$ & 
$1.40^{+0.02}_{-0.04}$ & $1.90^{+0.09}_{-0.05}$
\end{tabular}
\end{table*}

Inside the region accessible to Chandra,
the deprojected temperature and electron density
profiles for the X-ray gas
can be determined directly 
(under the assumption of spherical symmetry) using 
the methods described by
\citet{Allen01a} and \citet{Schmidt01}. These 
data are then used to calculate the 68 per
cent (1$\sigma$) confidence region of NFW mass models 
that provide the best-fit to the Chandra data.
The results on the NFW mass models for 
RX J1347.5-1145, Abell 1835 and Abell 478 are summarized in 
Tab.~\ref{tab:nfw}. 

We have carried out the extrapolation
procedure described in Sect.~\ref{extrap_proc} for every mass 
model within the 68\% confidence region obtained 
for each cluster. The radii
$r_1$ and $r_2$, the temperature $T_{r_1}$, the ratio $q$
(eq.~\ref{lokas}) calculated using the formulae in \citet{Lokas01}
and the effective power-law exponent $\gamma$ of the electron density
distribution for the clusters are given in Tab.~\ref{tab:extrap}. In
the case of Abell 478, the Chandra data do not extend beyond
$r\sim0.3h_{70}^{-1}$ Mpc (corresponding to an overdensity
$\Delta=8500$). Within this radius radiative cooling may have
affected the observed temperature of the cluster gas.
Following the instructions from Sect.~\ref{extrap_proc} for such a case,
the temperature profile and electron density profiles
were extrapolated from $r_{8500}$, but the exponent $\gamma$ was
determined using eq.~(\ref{lokas}) between $r_{2500}$ and $r_{200}$.

The application of our extrapolation procedure leads to 
the temperature and electron density profiles shown in
Fig.~\ref{profiles}. In this figure, both the best-fit profile and 
allowed range of models are shown. Note that the small scatter
in the electron density profiles is a consequence of the small scatter
in the luminosity profile derived from the Chandra data. The
temperature profiles have been binned using
a simple emission-weighting scheme.

For Abell 478, it is also 
possible to determine the power-law slope of the
electron density distribution beyond $r_1$ using ROSAT observations
\citep{Allen00}. Fitting the ROSAT data between $0.3$ Mpc and $0.9$
Mpc yields a slope of $\gamma=1.84\pm0.07$.
This result
is consistent with our model predictions given 
the expected steepening of the profile beyond the limit of the 
ROSAT data. 

Finally, we note that the power-law model used in extrapolating
the density profile provides only an approximation to the 
density values at large radii. (The true profile will steepen 
with increasing radius.) However, the pressure profile -- 
the relevant quantity for predicting the Comptonization parameter 
profile -- should be predicted accurately by the extrapolation 
procedure over the full range of radii.

\subsection{Combination with the SZ data}
\label{decrement}

\begin{table*}
\caption{Results: Column two lists the observed quantity used to measure
the Hubble constant. Columns three and four contain the 
measured and predicted values.
The fifth column lists the observed areas. Column six contains the
implied Hubble constant ($\Omega_{\rm m}=0.3$, $\Omega_{\Lambda}=0.7$).}
\label{results}
\begin{tabular}{llccll}
Object & quantity & observed & predicted & area 
& $H_{0}$\,[\kmpspMpc]\\
\hline
RX J1347.5-1145 & $\Delta F_{\nu}[{\rm mJy}]$ & -26.8$\pm2.4$ mJy &
$-25.5\pm2.3$ mJy  & 3
quadrants  & $63.4\pm16.1$ \\
Abell 1835 & SuZIE I scan & \multicolumn{2}{c}{direct fit}
 & 24 arcmin scan & $77.5\pm16.5$\\
Abell 478 & $\bar{y}\,[10^{-5}]$ & $7.52\pm0.56$ &
$7.43^{+0.12}_{-0.14}$ & 7.35' FWHM beam & $68.3\pm10.5$
\end{tabular}
\end{table*}

\subsubsection{RX J1347.5-1145}

Detections of the SZ effect in RX J1347.5-1145 were published
by~\citet[][P01]{Pointecouteau01} at 142.9 GHz, \citet{Komatsu01} at
21 GHz and 150 GHz and \cite[][R02]{Reese02} at
30GHz using the Diabolo, Nobeyama and Owens Valley Radio Observatory
(OVRO) instruments, respectively. The Comptonization parameters of all
groups are consistent with each other. We compare the Chandra
prediction with the Diabolo detection, which has the smaller beam size
at 142.9 GHz of the two bolometers (Diabolo and Nobeyama), as well as
a smaller error bar on the central Comptonization parameter than the
OVRO interferometer result. P01 also published the flux decrement of
the background radiation in four quadrants, which allows us to exclude
the south-east quadrant, which contains hot, probably
shocked, gas (Komatsu \etal 2001; Allen \etal 2002). 
Recently, \citet{Kitayama04} have also carried out a detailed study of
the intracluster medium in RX J1347.5-1145 using SZ observations. They
obtain results in good agreement with those determined from the
Chandra X-ray data.

In order to avoid the complicating 
effects of the shocked gas to the south-east of
the cluster centre in RXJ1347-1145, only data from the other three
quadrants were used to
determine the X-ray mass model \citep{Allen02}. P01 published the
flux decrements in the four quadrants, excluding a small $16.2$ arcsec
$\times$ 16.2 arcsec square region around the nucleus (to avoid
contamination by the central radio source).
Adding all four quadrants, they find flux decrement of $-(39.0\pm2.8)$
mJy in a square 100 arcmin $\times$ 100 arcmin area around the cluster
centre. However, this becomes $-(26.8\pm2.4)$ mJy once the south-east
quadrant is excluded.  This is the measurement we use for comparison
with the X-ray data.

The intensity change $\Delta I$ due to the SZ effect can be calculated
from the {Chandra} model. In the Kompaneets approximation
\citep[e.g., ][]{Birkinshaw99} this is given by
\begin{equation}
\Delta I=I_0\;y\;g(x)
\label{intensity}
\end{equation}
where $I_0=2(k_{\rm B}\,T_0)^3/h^2 c^2$,
\begin{equation}
g(x)=\frac{x^4 e^x}{\left(e^x-1\right)^2} \left[ \frac{x
    (e^x+1)}{e^x-1}-4
\right],
\end{equation}
and $x=h\nu/k_{\rm B}T_0$, with the Planck constant $h$, observing
frequency $\nu$ and microwave background temperature $T_0$. Where
relativistic corrections are significant, the Comptonization parameter
$y$ has to be modified appropriately (Sect.~\ref{step1}).

\subsubsection{Abell 1835}

The detection of the SZ effect in Abell 1835 was published
by~\citet[][M00]{Mauskopf00} using the Suzie I and Suzie II
bolometers. R02 have also published a detection using the OVRO
interferometer at 30 GHz, which provides a consistent, but more
precise determination of the Comptonization parameter. Here we 
compare our Chandra comptonization profiles with the co-added SuZIE I
scans published by M00, using their $\phi=-0.1$ arcsec offset of the
X-ray cluster centre from the scan centre. To calculate their difference
channels D3 (two beams separated by 4.6 arcmin) and T123 (triple beam
chop of three beams separated by 2.3 arcmin) of the bolometer array we
followed the description given by M00.

\subsubsection{Abell 478}

Abell 478 was observed by \citet{Myers97} at 32 GHz and reanalysed by
\citet[][MMR]{Mason01}. The observed decrement was published as an average
$y$-parameter $y=7.52\pm0.56\times10^{-5}$ within the telescope beam.
Note that we use their value without the relativistic correction, as
we apply this correction to the X-ray model. We compare the
observed SZ decrement with the {Chandra} prediction using the 
beam switching technique described by
\citet{Myers97}.

\section{Hubble constant determination}
\label{Hubble}

The Comptonization parameter $y$ defined in eq.~(\ref{y-integral})
depends upon the square root of the angular diameter distance to the
cluster \citep{Silk78,Cavaliere79}. By comparing the observed
Comptonization parameter, $y_{\rm obs}$, within a given beam or aperture
with the predicted value, $y_{\rm pred}$, from 
the X-ray data (for a given cosmology with $H_0=70\kmpspMpc$), 
we can measure the Hubble constant \eg (MMR), 

\begin{equation}
H_0=\left(\frac{y_{\rm pred}}{y_{\rm
obs}}\right)^2\times70\kmpspMpc.
\label{cosm1}
\end{equation}
Since eq.~(\ref{intensity}) is linear in the Comptonization parameter,
these considerations also apply to the flux decrement $\Delta
F_{\nu}=\omega \Delta I_{\nu}$ (Sect.~\ref{decrement}), where $\omega$
is the solid angle of the emitting area, and $\Delta I_{\nu}$ is the
intensity decrement.

In order to compare our X-ray-predicted Comptonization parameter
profiles with the direct SZ observations, we have convolved the X-ray
profiles with the instrument beams for RX J1347.5-1145 (22 arcsec FWHM
beam) and Abell 478 (7.35 arcmin FWHM beam and 22.16 arcmin beam
switching as described by \cite{Myers97}). For Abell 1835 we have
added an additional normalisation parameter to the NFW
parameter space of scale radius and concentration parameter
constrained by the Chandra data (Table~\ref{tab:nfw}). To determine the
best fit value and $1\sigma$ error for this normalisation parameter
we added the $\chi^2$ contribution from the SuZIE I co-added scans (1.7
arcmin FWHM beam and 2.3 arcmin or 4.6 arcmin beam separation)
to the $\chi^2$ contribution from the Chandra data. The
normalisation  parameter directly measures the distance scale and is
represented here by the resulting Hubble constant.

The results for the three galaxy clusters are shown in
Table~\ref{results}. 
For quantities with two-sided error bars, the
root-mean-square error was used to determine the error on the Hubble
constant. The weighted mean of the three Hubble constant
determinations in Table~\ref{results} is $H_0=69\pm8\kmpspMpc$.

\section{Comparison with the $\beta$-model and previous studies}

As the method for calculating the predicted Comptonization profiles 
presented here is significantly different from the standard
isothermal $\beta$-model approach, it is instructive to 
compare the results on $y(r)$ and the Hubble constant obtained with the two
approaches.

We have fitted the Chandra surface brightness profiles with 
a $\beta$-model (eq.~(\ref{betasurface})) inside a radius of 
2.5 arcmin for RX J1347.5-1145, 3 arcmin for Abell 1835, and between radii of 1.5 arcmin
and 4.5 arcmin for Abell 478. Note that in the case of Abell 478 the
core of the cluster had to be excluded, because the central surface
brightness profile is not flat, as is required by the
$\beta$-model. The mean, emission-weighted temperatures from 
single-temperature fits to the
Chandra spectra  and the $\beta$-model parameters are given in
Tab.~\ref{beta-fits}. The temperatures were determined
using the MEKAL \citep{Kaastra93,Liedahl95} plasma model. The gas density normalization $n_{{\rm e}\,0}$ was
calculated from the surface brightness using the MEKAL plasma emission
model with the spectroscopically determined
metallicities. Relativistic corrections were applied as described in
Sect.~\ref{step1}. Using eq.~(\ref{y0}), together with the appropriate 
relativistic
corrections factors, we obtain the central Comptonization parameters
$y_0=1.13\times10^{-3}$ for RX J1347.5-1145, 
$y_0=4.2\times10^{-4}$ for Abell 1835 and
$y_0=3.5\times 10^{-4}$ for Abell 478.

\begin{figure}
\resizebox{\columnwidth}{!}{\includegraphics{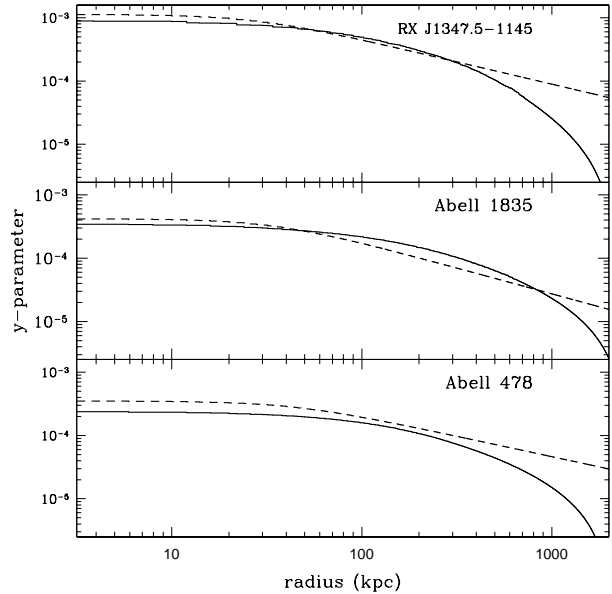}}
\caption{Best-fit fully deprojected and extrapolated {Chandra}
  Comptonization parameter profiles (solid lines). The results from 
$\beta$-model
  fits to the Chandra data are plotted with dashed lines.  All
  profiles are calculated for $H_0=70$\kmpspMpc, $\Omega_{\rm m}=0.3$
  and $\Omega_{\Lambda}=0.7$.}
\label{compton}
\end{figure}

Fig.~\ref{compton} shows the best-fit Comptonization parameter profiles
determined with our new method 
(solid line) together with the results from
the standard isothermal 
$\beta$-model approach (dashed line). This plot shows that the
Comptonization parameter profiles for the $\beta$-model differ
substantially from the profiles obtained using the new, 
deprojected/extrapolated solution. Firstly, the overall 
normalization in the core can
be quite different. Secondly, the slope of the $\beta$-model is too
shallow at large radii due to the infinite extent assumed in that model.
(We note that the $\beta$-model can yield significantly different 
answers when applied to other data sets covering different ranges of radii 
\citep{Pointecouteau01,Reese02,Mason01}. Here we have applied 
the two approaches to the same Chandra data simply to enable a 
direct comparison of the methods on the basis of the same, 
well-defined data sets.)

\begin{table}
\caption{$\beta$-model fits to the Chandra data}
\label{beta-fits}
\begin{tabular}{lcccc}
Object & $T$ & $\theta_{\rm c}$ & $n_{{\rm e}\,0}$ & $\beta$\\
& (keV) & (arcsec) & (cm$^{-3}$)\\
\hline
RX J1347.5-1145 & 12.2 & 4.8  & 0.207 & 0.57\\
Abell 1835      & 8.1  & 9.0  & 0.100 & 0.61\\
Abell 478       & 6.8  & 26.6 & 0.068 & 0.55
\end{tabular}
\end{table}

Applying the procedure described in Sect.~\ref{Hubble} to 
the measured $\beta$-profile parameters 
yields Hubble constant estimates 
of $H_0=32\kmpspMpc$ for RX J1347.5-1145, $H_0=29\kmpspMpc$ for
Abell 1835 and $H_0=103\kmpspMpc$ for Abell 478. One can 
immediately see that the application of the isothermal 
$\beta$-model does not yield a consistent 
estimate when applied naively to Chandra data for the present
sample of clusters.

It is also interesting to compare these estimates with
previously published $\beta$-model estimates for the clusters
discussed here. We have translated the electron densities from 
other studies into our assumed cosmology ($\Omega_{\rm m}=0.3$,
$\Omega_{\Lambda}=0.7$ and $H_0=70\kmpspMpc$).

\begin{enumerate}
\item[{\bf RX J1347.5-1145:}] P01 determined a Hubble constant
  $H_0=44\pm6\kmpspMpc$ using a ${\rm k}T=9.3$ keV isothermal
  model with the $\beta$-model parameters $\theta_{\rm c}=8.4$
  arcsec, $\beta=0.56$ and $n_{{\rm e}\,0}=0.102\,$cm$^{-3}$ from ROSAT
  \citep{Schindler97}.
  Using the same temperature and a slightly different $\beta$-model,
  R02 found an angular diameter distance that implies
  $H_0=68.2^{+26.7}_{-15.8}\kmpspMpc$. This estimate 
  is consistent with the value we obtain using our fully deprojected
  and extrapolated cluster model (Tab.~\ref{results}). However, the
  gas temperature and the other $\beta$-model parameters used by R02
  are not consistent with the $\beta$-model parameters from Chandra
  (Tab.~\ref{beta-fits}).

\item[{\bf Abell 1835:}] For this cluster, M00 measured a
Hubble constant $H_0=66^{+38}_{-22}\kmpspMpc$ for a ${\rm k}T=9.8$ keV
isothermal model with the $\beta$-model parameters $\theta_{\rm
c}=13.2$ arcsec, $\beta=0.58$ and $n_{{\rm e}\,0}=0.058\,$cm$^{-3}$ from
a fit to ROSAT data. This temperature was obtained by allowing for a
cooling flow component in the core of the cluster. R02 obtained
an angular diameter distance to the cluster which implies
$H_0=55.4^{+13.2}_{-8.8}\kmpspMpc$, based on a ${\rm k}T=8.21$ keV
isothermal model (no cooling flow correction). The difference between
these two measurements can be attributed to the different isothermal
temperatures and small differences in the assumed $\beta$-model
parameters. Note also that the $\beta$-model parameters used in these 
studies are
significantly different from those determined from Chandra
(Tab.~\ref{beta-fits}) over a smaller range of radii. 
This explains the difference in the predicted
Hubble constant. However, the M00 result is consistent with the result we
obtain from the deprojected and extrapolated X-ray data, which shows
that it is possible to partially correct the $\beta$-model in the
central region of the cluster when the surface brightness profile
observed with the ROSAT field of view is used. For larger radii this
approximation will fail, however, because of the fixed $\beta$-model slope.

\item[{\bf Abell 478:}] MMR measured a Hubble constant
$H_0=61^{+33}_{-21}\kmpspMpc$ assuming an isothermal $kT=8.4$ keV
model with the $\beta$-model parameters $\theta_{\rm c}=1$ arcmin,
$\beta=0.64$ and $n_{{\rm e}\,0}=0.023\,$cm$^{-3}$ from ROSAT. This
temperature also accounts for the presence of a cooling flow in the
cluster core.
\citet{Sun03} improved upon the result by MMR by determining the
Comptonization parameter as a function of radius from a deprojection
of the Chandra data, and by extrapolating the Chandra data with a
$\beta$-model based on a fit to the combined surface brightness
profile from Chandra and ROSAT. They used $\beta=0.68$ and obtained a
Hubble constant estimate $H_0=64^{+32}_{-18}\kmpspMpc$, which lies
between our value and the one obtained by MMR. 

\end{enumerate}

\section{Summary and discussion}
\label{conclusions}

We have presented a new method for extrapolating the pressure profiles of
galaxy clusters beyond the region accessible to current X-ray
satellites such as Chandra and XMM-Newton, for use in 
combined X-ray and SZ studies. Our method assumes hydrostatic equilibrium, 
an NFW mass profile for the clusters and predicts the temperature profile
and an effective power-law
slope of the electron density profile.
The method does not assume isothermality and provides a simple
extrapolation recipe that can be easily implemented. We have applied
our method to the Chandra data for three X-ray bright, dynamically
relaxed galaxy clusters, RX J1347.5-1145, Abell 1835 and Abell 478, and
have obtained detailed temperature, electron density and
Comptonization parameter profiles for these clusters out to the
virial radius.

Since all three clusters have published SZ detections, we were also
able to measure the Hubble constant for each, as shown in
Table~\ref{results}. The individual 
measurements are consistent with each other
and yield a weighted mean of $H_0=69\pm8$\kmpspMpc.
This result is consistent with findings from the 
Hubble Key Project ($H_0=72\pm8$\kmpspMpc; \citealt{Freedman01})
and combined studies of the cosmic microwave background
and X-ray galaxy cluster data ($H_0=68.4^{+2.0}_{-1.4}$ \kmpspMpc;
\citealt*{Allen03}). Note that our weighted mean value 
does not include
additional systematical errors, due to, e.g., clumping or asphericity,  
although for such regular, relaxed
clusters these should be at a minimum.

The contribution to the predicted Comptonization parameter from
material beyond the regions directly observed with Chandra, 
modelled by our extrapolation procedure, varies from system to 
system. For RX J1347.5-1145 the correction to the Comptonization
parameter in the region used in Table~\ref{results} amounts only 
to $1$ per cent and is negligible. For Abell 1835, however, 
the contribution from material
in the extrapolated region amounts to $20$ per cent, and for Abell 478
the contribution rises to $68$ per cent. This shows that a robust
extrapolation method is necessary when combining X-ray and SZ
observations to infer cosmological information.

We have estimated the effect of changing the outer radius $r_{200}$ by
repeating the calculation for Abell 478, but stopping at
$r_{500}$. This leads to a 4.2 per cent smaller average
Comptonization parameter, well within the observational error
bar. 

We have compared the Comptonization parameter profiles predicted by
our method with the profiles obtained by fitting an isothermal 
$\beta$-model to
the Chandra data. We find that a naive application of 
the $\beta$-model approach 
leads to profiles with 
significantly different central Comptonization parameters
and predicts different slopes for the Comptonization parameter
profiles at large radii. The isothermal $\beta$-model also leads to 
inconsistent Hubble constant estimates 
when applied to the Chandra data sets considered
here. 

It is clear that for galaxy clusters like Abell 478, a
field of view larger than the one afforded by the Chandra ACIS-S3
detector would be beneficial. In this respect the combination of
Chandra and XMM-Newton observations will help (\eg Pointecouteau \etal 2004), 
although the detector background will still 
prohibit precise measurements of the
temperature profile beyond $r\sim 0.5r_{200}$. (We also note
that the high spatial resolution of Chandra is important 
in resolving
the central temperature profile and constraining the best-fit NFW mass
models.)

Observations for a larger sample of clusters, 
as well as deeper X-ray and SZ observations, should make this a
powerful method to explore the cosmological distance scale. 
In the first case, it will be important to
concentrate such studies on the largest, 
dynamically relaxed clusters 
for which systematic uncertainties associated with the 
deprojection method and assumption of hydrostatic 
equilibrium are at a minimum.

\section*{Acknowledgments}

We thank Y. Suto, N. Itoh and the anonymous referee for comments.
We thank the developers of the GNU Octave numerical computation
language and the GNU Scientific Library for these tools. SWA and ACF
acknowledge the support of the Royal Society.

\appendix

\section{An extrapolation solution for the X-ray gas temperature profile in an
  NFW potential}
\label{extrap}

Let the mass distribution of a galaxy cluster be parametrized by a
NFW mass model (eq.~\ref{nfw}). Also, let the electron density
distribution for a certain range of radii be parametrized by a
power-law model
\begin{equation}
n_{\rm e}(r)\sim r^{-\gamma}.
\label{nepow}
\end{equation}
The temperature as a function of radius can then be determined under
the assumption of hydrostatic equilibrium by solving the hydrostatic
equation
\begin{equation}
\frac{1}{n_{\rm e}}\frac{{\rm d}(n_{\rm e}\;kT)}{{\rm d}r} = - \mu
m_{\rm p} \frac{{\rm G}M}{r^2},
\label{hydro}
\end{equation}
where $k$ is the Boltzmann constant, $\mu$ is the molecular weight of
the gas and $m_{\rm p}$ is the proton mass (note that the
normalization of (\ref{nepow}) drops out of this equation). Using the
method of the variation of the constant one finds that is possible to
write the solution of (\ref{hydro}) in the form
\begin{eqnarray}
kT(r) & =& \left(\frac{r}{r_1} \right)^{\gamma} \left(k T_1- \frac{4 \pi
r_{\rm s}^2 \delta_{\rm c} \rho_{\rm c} \mu m_{\rm p}
{\rm G}}{(1+\gamma)\, r_1 r^{1+\gamma}}\times \right.
\nonumber\\
&& \left( r\, r_1^{1+\gamma}\,
{}_2F_1(-\gamma,1;1-\gamma;-\frac{r}{r_{\rm s}})\right.\nonumber\\
&&-r_1\,r^{1+\gamma}\,
{}_2F_1(-\gamma,1;1-\gamma;-\frac{r_1}{r_{\rm s}}) \nonumber\\
&& \left.\left.- r_{\rm s}
r_1^{1+\gamma}\, {\rm ln}\!\left(1+\frac{r}{r_{\rm s}} \right)
+r_{\rm s} r^{1+\gamma}\, {\rm ln}\!\left(1+\frac{r_1}{r_{\rm s}}
\right)\right) \right)\!\!,
\label{tprof_analyt}
\end{eqnarray}
with the boundary condition $T(r_1)=T_1$. ${}_2F_1$ is the
hypergeometric function \citep[e.g., ][ p. 995]{Gradshteyn00}
which is well-defined on the whole negative real axis and is available
in standard mathematical packages.

\section{The energy distribution of the NFW mass distribution}
\label{lokmam}

In this Appendix we list the \citet{Lokas01} expressions for the
potential energy $W_{\rm pot}$ and the kinetic energy $W_{\rm kin}$ in
an NFW potential with an isotropic velocity dispersion, as a function of
radius $r$. The ratio of these quantities is used in
Sect.~\ref{extrap_proc} to determine the cluster model extrapolation.
It can be calculated from the concentration parameter $c$ and the
scale radius $r_{\rm s}$.

The radial dependence of the potential energy associated with an NFW
mass distribution with virial radius $r_{200}=c\times r_{\rm s}$ as a
function of (scaled) radius $s=\frac{r}{r_{200}}$ is
\begin{eqnarray}
    W_{\rm pot}(s) 
    &=& - W_\infty \left[ 1-\frac{1}{(1+c s)^2}
    - \frac{2 \ln (1+c s)}{1+c s} \right]
\end{eqnarray}
(eq.~21 in \citealt{Lokas01}), where $W_\infty$ is the asymptotic value
\begin{equation}   
    W_\infty =
    \frac{G M_{\rm tot}^2(r_{200})}{2 r_{\rm s} [{\rm ln}(1+c)-{c/(1+c)}]^2} .
\end{equation}
The radial dependence of the kinetic energy, assuming isotropic
velocity dispersion, is
\begin{eqnarray}
    W_{\rm kin}(s) &=&
    \frac{1}{2} W_\infty \{ -3 + \frac{3}{1+c s} - 2 \ln (1+c s)
    \nonumber \\
    && +  c s [5 + 3 \ln (1+ c s)]
    - c^2 s^2 [7 + 6 \ln (1+ c s)]
    \nonumber \\
    && +  c^3 s^3 [\pi^2 - \ln c - \ln s +
    \ln (1+c s) \nonumber \\
    && +   3 \ln^2 (1+c s) +
    6 {\rm Li}_2 (-c s)] \}
\end{eqnarray}
(eq.~24 in \citealt{Lokas01}), where Li$_2$ is the dilogarithm.

\bsp

\label{lastpage}


\begin{thebibliography}{99}
\bibitem[\protect\citeauthoryear{Allen}{2000}]{Allen00} Allen S.W.,
  2000, MNRAS, 315, 269
\bibitem[\protect\citeauthoryear{Allen, Ettori \& Fabian}{Allen et
al.}{2001}]{Allen01a} Allen S.W., Ettori S., Fabian A.C., 2001, MNRAS,
324, 877
\bibitem[\protect\citeauthoryear{Allen, Schmidt \& Fabian}{Allen et
al.}{2001}]{Allen01b} Allen S.W., Schmidt R.W., Fabian A.C., 2001,
MNRAS, 328, L37
\bibitem[\protect\citeauthoryear{Allen, Schmidt \& Fabian}{Allen et
al.}{2002}]{Allen02} Allen S.W., Schmidt R.W., Fabian A.C., 2002,
MNRAS, 335, 256
\bibitem[\protect\citeauthoryear{Allen, Schmidt \& Bridle}{Allen et
al.}{2003}]{Allen03} Allen S.W., Schmidt R.W., Bridle S.L., 2003,
MNRAS, in press
\bibitem[\protect\citeauthoryear{Birkinshaw}{1999}]{Birkinshaw99}
Birkinshaw M., 1999, Phys. Rep., 310, 97
\bibitem[\protect\citeauthoryear{Carlstrom et al.}{2001}]{Carlstrom01}
Carlstrom J.E., Joy M., Grego L., Holder G., Holzapfel W.L., LaRoque
S., Mohr J.J., Reese E.D., to appear in: ``Constructing the Universe
with Clusters of Galaxies'', IAP conference, July 2000, eds. F. Durret
and G. Gerbal, preprint: astro-ph/0103480
\bibitem[\protect\citeauthoryear{Cavaliere \&
Fusco-Femiano}{1978}]{Cavaliere78} Cavaliere A., Fusco-Femiano R.,
1978, A\&A, 70, 677
\bibitem[\protect\citeauthoryear{Cavaliere, Danese, \& de
Zotti}{Cavaliere et al.}{1979}]{Cavaliere79} Cavaliere A., Danese
L., de Zotti G., 1979, A\&A, 75, 322
\bibitem[\protect\citeauthoryear{Challinor \&
Lasenby}{1998}]{Challinor98} Challinor A., Lasenby A., 1998, ApJ, 499,
1
\bibitem[\protect\citeauthoryear{Eke, Navarro, \& Frenk}{Eke et
al.}{1998}]{Eke98} Eke V.R., Navarro J.F., Frenk C.S., 1998, ApJ, 503,
569
\bibitem[\protect\citeauthoryear{Freedman et al.}{2001}]{Freedman01}
Freedman W.L., et al., 2001, ApJ, 553, 47
\bibitem[\protect\citeauthoryear{Gradshteyn \& Ryzhik}{2000}]{Gradshteyn00}
Gradshteyn I.S., Ryzhik I.M., 2000, ``Table of Integrals, Series and
Products'', Sixth Edition, Academic Press, San Diego
\bibitem[\protect\citeauthoryear{Inagaki, Suginohara, \& Suto}{Inagaki
    et al.}{1995}]{Inagaki95} Inagaki Y., Suginohara T., Suto, Y.,
    1995, PASJ, 47, 411
\bibitem[\protect\citeauthoryear{Kaastra et al.}{2004}]{Kaastra04} 
Kaastra J.S., et al., 2004, A\&A, 413, 415
\bibitem[\protect\citeauthoryear{Itoh et al.}{Itoh, Kohyama \&
Nozawa}{1998}]{Itoh98} Itoh N., Kohyama Y., Nozawa S., 1998, ApJ, 502,
7
\bibitem[\protect\citeauthoryear{Jones et al.}{2001}]{Jones01} Jones
M.E., et al., 2001, submitted to MNRAS, preprint: astro-ph/0103046
\bibitem[\protect\citeauthoryear{Kaastra \& Mewe}{1993}]{Kaastra93} Kaastra
J.S., Mewe R., 1993, Legacy, 3, 16
\bibitem[\protect\citeauthoryear{Kitayama et
al.}{2004}]{Kitayama04} Kitayama T., et al., 2004, PASJ, 56, 17
\bibitem[\protect\citeauthoryear{Komatsu et al.}{2001}]{Komatsu01}
Komatsu E., et al., 2001, PASJ, 53, 57
\bibitem[\protect\citeauthoryear{Liedahl et al.}{1995}]{Liedahl95} Liedahl
D.A., Osterheld A.L., Goldstein W.H., 1995, ApJ, 438, L115
\bibitem[\protect\citeauthoryear{Lokas \& Mamon}{2001}]{Lokas01}
  Lokas E.L., Mamon G.A., 2001, MNRAS, 321, 155
\bibitem[\protect\citeauthoryear{Mason, Myers \& Readhead}{Mason et
    al.}{2001}]{Mason01} Mason B.S., Myers S.T., Readhead A.C.S, 2001,
    ApJL, 555, 11 (MMR)
\bibitem[\protect\citeauthoryear{Mauskopf et al.}{2000}]{Mauskopf00}
Mauskopf P.D., et al., 2000, ApJ, 538, 505 (M00)
\bibitem[\protect\citeauthoryear{Myers et al.}{1997}]{Myers97} Myers
  S.T., Baker J.E., Readhead A.C.S., Leitch E.M., Herbig T., 1997,
  ApJ, 485, 1
\bibitem[\protect\citeauthoryear{Navarro, Frenk \& White}{Navarro et
al.}{1995}]{Navarro95} Navarro J.F., Frenk C.S., White S.D.M., 1995,
MNRAS, 275, 720 (NFW)
\bibitem[\protect\citeauthoryear{Pointecouteau et
al.}{2001}]{Pointecouteau01} Pointecouteau E., Giard M., Benoit A.,
D{\'e}sert F.X., Bernard J.P., Coron N., Lamarre J.M., 2001, ApJ, 552,
42 (P01)
\bibitem[\protect\citeauthoryear{Pointecouteau et
al.}{2004}]{Pointecouteau04} Pointecouteau E., Arnaud M., Kaastra J., 
de Plaa J., 2004, A\&A, submitted (P04)
\bibitem[\protect\citeauthoryear{Reese et al.}{2002}]{Reese02}
  Reese E.D., Carlstrom J.E., Joy M., Mohr J.J., Grego L., Holzapfel
  W.L., 2002, ApJ 581, 53 (R02)
\bibitem[\protect\citeauthoryear{Sarazin}{1988}]{Sarazin88} Sarazin
C.L., 1988, ``X-ray emission from clusters of galaxies'', Cambridge
University Press, Cambridge
\bibitem[\protect\citeauthoryear{Sazonov \& Sunyaev}{1998}]{Sazonov98}
Sazonov S.Y., Sunyaev R.A., 1998, ApJ, 508, 1
\bibitem[\protect\citeauthoryear{Schindler et al.}{1997}]{Schindler97}
Schindler S., Hattori M., Neumann D.M., Boehringer H., 1997,
A\&A, 317, 646
\bibitem[\protect\citeauthoryear{Schmidt, Allen \& Fabian}{Schmidt et
al.}{2001}]{Schmidt01} Schmidt R.W, Allen S.W., Fabian A.C., 2001,
MNRAS, 327, 1057
\bibitem[\protect\citeauthoryear{Silk \& White}{1978}]{Silk78} Silk
J., White S.D.M., 1978, ApJL, 226, 103
\bibitem[\protect\citeauthoryear{Sun et al.}{2003}]{Sun03} Sun M.,
  Jones C., Murray S.S., Allen S.W., Fabian A.C., Edge A.C., 2003,
  ApJ, 587, 619
\bibitem[\protect\citeauthoryear{Sunyaev \&
Zeldovich}{1972}]{Sunyaev72} Sunyaev R., Zeldovich Y., 1972, Comments
Astrophys. Space Phys., 4, 173
\bibitem[\protect\citeauthoryear{Yoshikawa, Itoh \& Suto}{Yoshikawa
    et al.}{1998}]{Yoshikawa98} Yoshikawa K., Itoh M., Suto Y., 1998,
    PASJ, 50, 203
\end{thebibliography}
\end{document}